\documentclass[12pt]{article}
\pdfoutput=1
\usepackage{jheppub}
\usepackage{verbatim}
\usepackage{amsmath}
\usepackage{amssymb}
\usepackage{graphicx}
\usepackage{amsthm}
\usepackage{slashed}
\usepackage{amsfonts}
\usepackage[utf8]{inputenc}
\usepackage[normalem]{ulem}
\usepackage{tikz}
\usepackage{dsfont}
\usepackage{braket}
\usepackage{mathrsfs}

\usepackage{array}

\usepackage{float}

\def\be{\begin{equation}}
\def\ee{\end{equation}}
\def\ba#1\ea{\begin{align}#1\end{align}}
\def\bg#1\eg{\begin{gather}#1\end{gather}}
\def\bm#1\em{\begin{multline}#1\end{multline}}
\def\bmd#1\emd{\begin{multlined}#1\end{multlined}}

\def\dd{\text{d}}

\def\({\left(}
\def\){\right)}
\def\[{\left[}
\def\]{\right]}
\def\<{\langle}
\def\>{\rangle}

\newcommand{\bfig}{\begin{figure}\begin{center}}
\newcommand{\efig}{\end{center}\end{figure}}
\newcommand{\bi}{\begin{itemize}}
\newcommand{\ei}{\end{itemize}}

\theoremstyle{definition}

\begin{document}
\begin{flushright}
IFT-UAM/CSIC-19-165
\end{flushright}
\title{\LARGE Momentum/Complexity Duality \\ and the Black Hole Interior}
\author{ Jos\'e~L.~F. Barb\'on,~Javier~Mart\'in-Garc\'ia and ~Martin~Sasieta}
\affiliation{Instituto de Física Teórica, IFT-UAM/CSIC \\  c/ Nicolás Cabrera 13, Universidad Autónoma de Madrid, 28049, Madrid, Spain}
\emailAdd{jose.barbon@csic.es, javier.martingarcia1@gmail.com, martin.sasieta@csic.es}

\abstract{We establish a version of the Momentum/Complexity (PC) duality  between the rate of operator complexity growth and an appropriately defined radial component of bulk momentum for a test system falling into a black hole. In systems of finite entropy, our map remains  valid for arbitrarily late times after  scrambling. The asymptotic regime of linear complexity growth is associated to a frozen momentum in the interior of the black hole, measured with respect to a time foliation by extremal codimension-one surfaces which saturate without reaching the singularity.  The detailed analysis in this paper uses the Volume-Complexity (VC) prescription and  an infalling system consisting of a thin shell of dust, but the final PC duality formula should have a much wider degree of generality. 
}

\maketitle

\section{Introduction}
\noindent

Measures of operator complexity have received  considerable recent attention in studies of  information scrambling in
many-body quantum systems \cite{RobertsStanford, QiStreicher, HuseHydro, HuseLyapunov, Altman, BarbonRitam, Mousatov}. One  motivation is the
characterization of quantum complexity in holographic systems.   In that  context, it has been proposed that the `size' of an operator can be characterized by a mechanical momentum of an effective particle in the bulk (cf. \cite{SusskindThingsFall,  Magan, SusskindNewton, SusskindSchwarzian}). The bulk particle is `injected' by the `small'  operator ${\cal O}$ on the boundary, acting on some reference state $
{\cal O} \,|\Psi\rangle $
at, say $t=0$.  If the resulting state is evolved in time 
\be\label{teinject}
e^{-it H} \,{\cal O} \,|\Psi\rangle =e^{-itH} \,{\cal O} \,e^{itH} \,e^{-itH} \,|\Psi\rangle =  {\cal O}_{-t} \,|\Psi\rangle_t 
\;,
\ee
any increase of complexity can be attributed partly to the increase in complexity of the time-evolved reference state $|\Psi\rangle_t$, and partly to the increase in complexity of the operator when evolved to the past, in what we usually refer to a `precursor': ${\cal O}_{-t} = e^{-itH} {\cal O}\,e^{itH}$. If the increase in complexity of the reference state can be neglected or somehow subtracted, we can define the complexity of the operator ${\cal O}_{-t}$ in terms of the complexity of the evolved state.  The state  (\ref{teinject})  can be interpreted as a heavy particle state falling through the bulk. More precisely, we may define the operator complexity in terms of the state complexity by the subtraction
\be\label{opsub}
{\cal C}_{\cal O} (t) = {\cal C}\left[{\cal O}_{-t} |\Psi\rangle_t \right] - {\cal C}\left[ |\Psi\rangle_t\right]\;,
\ee
with some appropriate normalization. 
In practice, this definition must be supplemented by some definite prescription for the state complexity such as, for example, the AC/VC definitions (cf. \cite{SusskindStanford1,SusskindZhao, SusskindShocks,SusskindNotEnough, SusskindACShort, SusskindACLong}).

Let us suppose that the state  (\ref{teinject})  can be interpreted as a heavy particle  falling through the bulk. Then, the momentum/complexity duality proposal (PC duality for short)   amounts to a relation of the form
\be\label{mcd}
{\dd{\cal C}_{\cal O} \over \dd t} = P_{\cal C}\;,
\ee
where ${\cal C}_{\cal O}$ is the complexity of the operator, and $P_{\cal C}$ is a suitable  component of the mechanical momentum of the associated particle. On general grounds, the right-hand side of (\ref{mcd}) has an inherent ambiguity, since we must specify which particular momentum component is the relevant one, and this selects a particular coordinate system. 
A simple example which illustrates this fact is obtained by regarding the free fall of a particle in a Rindler near-horizon region as dual to
operator growth in a fast scrambler. In the vicinity of a regular horizon we can pick polar Rindler coordinates $(\rho, t)$  which approximate the metric as
\be\label{rin}
\dd s^2 \approx -\kappa^2 \rho^2 \dd t^2 +  \dd \rho^2 +  \dd s^2_\perp\;,
\ee
where $ \dd s^2_\perp$ is a metric along the horizon which formally sits at $\rho=0$, and $\kappa$ is the surface gravity. A particle  with action 
$$
S_P  =m\int  \dd t \,L_P =  -m \int  \dd t\, \sqrt{\kappa^2 \rho^2 - \left({ \dd \rho \over  \dd t}\right)^2 + \dots} 
$$
falling towards the horizon along any causal path follows the law $\rho \approx \rho_0 \,\exp(-\kappa t)$ at late Rindler times,  and the Rindler-radial momentum satisfies 
\be\label{radm}
P_\rho = {\partial L_P \over \partial {\dot \rho}} \propto e^{\kappa t} \;,
\ee
where ${\dot \rho \equiv  \dd \rho / \dd t}$. Since the surface gravity coincides with the fast-scrambling Lyapunov exponent, $\kappa = \lambda_L$, the idea is to relate $P_\rho$ and operator size. 
 In this case, both terms in (\ref{mcd}) grow exponentially in time, so that the qualitative behavior only establishes $P_{\cal C} \sim P_\rho$ as proportional to the complexity, or any of its higher time derivatives. A more precise matching can be obtained by testing the PC duality in near-extremal Reissner--Nordstrom horizons. In this case, there is a `pre-scrambling' period corresponding to the fall through the AdS$_2$ throat which, upon comparison with detailed calculations of operator growth in the SYK model \cite{QiStreicher, RN, SusskindNewton}, leads to (\ref{mcd}).

\begin{figure}[h]
	\centering
	\includegraphics[width = .5\textwidth]{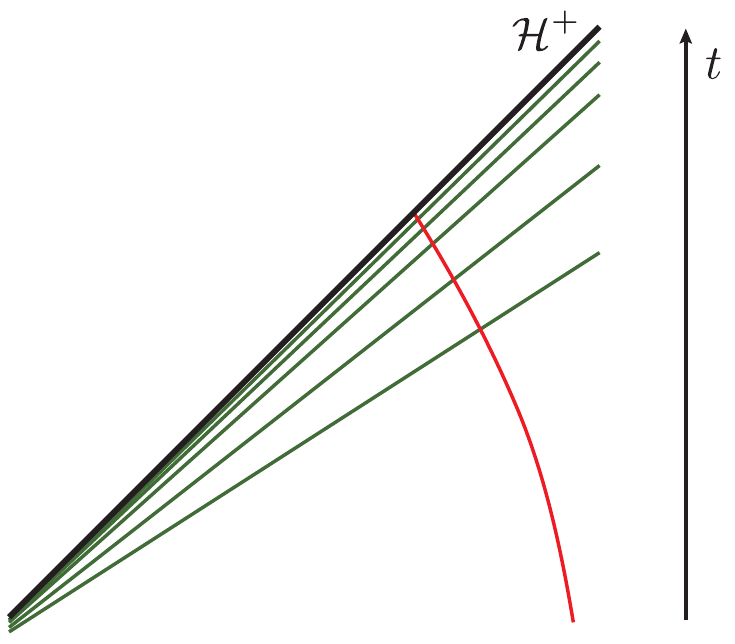}
	\caption{Standard notions of PC duality are defined in terms of near-horizon dynamics, using radial and time coordinates which remain outside the horizon. }
	\label{susskindfoliation}
\end{figure}

Such notions of PC duality  involve the particle fall towards the horizon, as indicated in Figure \ref{susskindfoliation},  and an interpretation in terms of operator size in the quantum mechanical dual system. In systems with finite size, operator growth as such should stop at the scrambling time, of order $t_s \sim \lambda_L^{-1} \,\log \,N_{\rm eff}$, where $N_{\rm eff}$ is the effective number of degrees of freedom. In the picture of bulk infall the scrambling time corresponds to the particle reaching the stretched horizon, a timelike layer situated about one Planck length away from the horizon. 

An interesting question is whether it is possible to establish a different type of PC correspondence for operator complexity that would operate at times much larger than the scrambling time. In this regime, complexity and size are not expected to be proportional: while operator size should saturate, an operator complexity defined as in (\ref{opsub}) should grow linearly at long times, with a slope proportional to the average energy injected in the system by the action of the operator. This is expected in tensor-network or quantum circuit definitions of complexity, but it also seems to hold  in different definitions of operator complexity, such as K-complexity (cf.  \cite{Altman}), which  was recently shown to  exhibit the characteristic  linear growth at  late times (cf. \cite{BarbonRitam}). 

It is natural to expect that any form of  operator PC correspondence that accesses the late time linear regime would depend on kinematical properties of trajectories in the black hole interior. If this is so, it is interesting to learn what those concrete properties would be.    In this paper we show that, adopting complexity=volume prescription (VC) as the definition of  (\ref{opsub}), a PC correspondence of the form (\ref{mcd})  exists at all times,  for operators that are dual to spherical shells falling on timelike trajectories. The momentum $P_{\cal C}$ is that of the shells, measured with respect to a particular radial coordinate which we specify.  More precisely, we find
\be\label{giveaway}
{\dd{\cal C}_{\cal O} \over \dd t} = P_{\cal C} (t) = -\int_{\Sigma_t} N^\mu_\Sigma \,T_{\mu\nu}\,{\cal C}^\nu_\Sigma \;,
\ee
where $\Sigma_t$ is a maximal-volume surface anchored at boundary time $t$, the basic ingredient of the VC definition, $N_\Sigma$ is the unit normal to $\Sigma_t$ and ${\cal C}_\Sigma$ is a suitable radial vector field defined on $\Sigma_t$. 
 In this form of the PC correspondence, the shells only contribute through their energy momentum tensor, and the `suitable coordinate system' to measure the momentum is obtained by foliating the bulk spacetime with the extremal-volume surfaces themselves. Therefore, we expect  (\ref{giveaway}) to have a much wider generality than the thin-shell dynamics which was used for its derivation.   The compatibility of a constant late-time complexity rate   and a  constant bulk matter momentum results form the late-time accumulation of maximal surfaces in the black hole interior, a well-known property of the VC prescription.

The paper is organized as follows: in section \ref{sec:collapse} we describe the class of operators for which we establish the PC duality. In section \ref{sec::proof} we give a proof of (\ref{giveaway}) in this context. We end with conclusions and three appendices containing generalizations and some technical points.

\section{Thin-shell operators and states}
\label{sec:collapse}
\noindent

For a holographic CFT defined on a spherical spatial manifold ${\bf S}^{d-1}$ of  radius $L$, we consider its gravity dual on AdS$_{d+1}$, also taken to have  curvature radius $L$.   A thin shell of dust injected from the  AdS boundary can be represented in the CFT by the action of a  formal product operator 
\be\label{forp}
{\cal O}_{\rm shell} \sim \prod_{D_\Lambda \in {\cal P}_\Lambda} \phi_{\Lambda, {D_\Lambda}} \; ,
\ee
where ${\cal P}_\Lambda$ is a partition of the sphere in domains $D_\Lambda$ of size $\Lambda^{-1}$, the regularization cutoff. The operators $\phi_{\Lambda, D_{\Lambda}}$ can be seen as  bulk operators, applied at radius of order $r_\Lambda \sim \Lambda \,L^2$, and smeared over the domain $D_\Lambda$.  The idea is to use $\phi_{\Lambda, D_\Lambda}$ to inject a heavy bulk particle at radius $r_\Lambda$. Although we imagine specifying the operators in bulk effective field theory, we can always regard it as a CFT operator by a bulk-boundary reconstruction map, say using the HKKL formulation \cite{HKLL}.

These operators are `big' in the sense of the spatial structure, but are `simple' in holographic terms, since they are constructed from operators near the boundary of AdS. By appropriately choosing $\phi_{\Lambda, D_\Lambda}$, we can generate a semiclassical  state whose subsequent evolution is parametrized as the collapse of the shell of particles in the bulk geometry. In the case that the local factors $\phi_{\Lambda, D_\Lambda}$ are engineered with very massive bulk fields, or equivalently CFT operators with very large conformal weight, we can regard the shell as composed of classical massive particles forming a dust cloud with density $\sigma$ and four-velocity field $u^\mu$. 

For the purposes of this paper, we define the operator complexity in terms of the general prescription (\ref{opsub}), where the state complexity is regarded as computed with the VC prescription. For technical convenience, we shall take the high-temperature thermofield double state as the reference state on the Hilbert space of two copies of the CFT, and the shell state is injected on the Right CFT as indicated in Figure \ref{susskindfoliation}, at times much larger than the thermalization time $T^{-1}$, where $T$ is the Hawking temperature of the black hole. 

\begin{figure}[h]
\centering
\includegraphics[width = 0.6\textwidth]{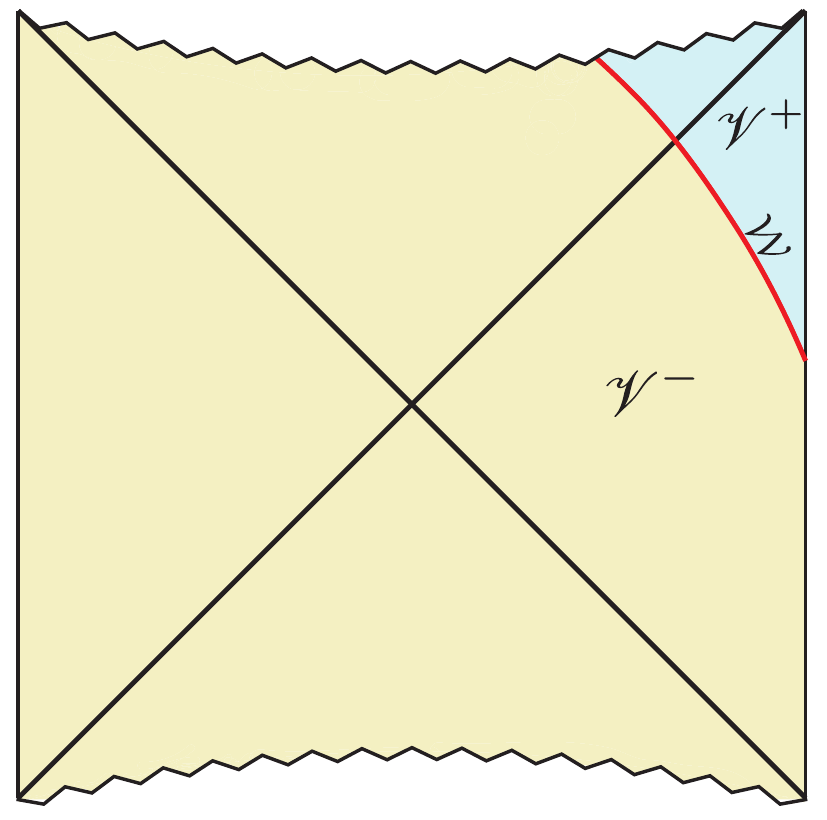}
\caption{Penrose diagram of the collapsing shell geometry. The shell is injected in the bulk at late times compared with $ T^{-1}$, causing the initial black hole of mass $M_-$ to grow up to the bigger mass $M_+$. The worldvolume of the matter shell is labelled $\mathcal{W}$ and sets the boundary between the two black hole spacetimes $\mathscr{V}^\pm$. }
\label{penrose}
\end{figure}

The complexity of the shell operator is  defined in terms of bulk quantities  as 
\begin{equation}
\label{cdef}
{\cal C}\left[{\cal O}_{\rm shell}\right] = {d-1 \over 8\pi\, G \,L} \left[ {\rm Vol} (\Sigma_{\rm bh+shell}) - {\rm Vol}(\Sigma_{\rm bh})\right]\;,
\end{equation}
where $\Sigma$ denotes the extremal codimension-one hypersurface with given asymptotic boundary conditions, defined in the eternal black hole spacetime with and without the shell. The concrete prefactor in (\ref{cdef}) is chosen for convenience of normalization.  From now on shall measure bulk lengths in units of curvature radius, so that we set $L=1$. 

The worldvolume of the thin shell is a codimension-one timelike manifold ${\cal W}$ which divides the spacetime manifold in two regions:  $\mathscr{V^+}$ is a Schwarzschild-AdS solution of mass $M_+$ which we identify as `exterior' or `right' region, and $\mathscr{V^-}$, a similar solution of mass $M_-$ referred to as the `interior'  or `left' region. The ADM energy of the shell is given by $M_+ - M_-$ and is assumed to be positive. Spherical symmetry holds  globally in the full spacetime, whereas stationarity is broken at ${\cal W}$. Both $\mathscr{V^\pm}$ have smooth Killing vectors which are timelike in the asymptotic regions  and spacelike inside event horizons. Denoting these vectors as  $\xi_\pm = \partial/\partial t_\pm$, where $t_\pm$ are adapted coordinates, we can write a standard form of the metric  on both sides of ${\cal W}$: 

\begin{eqnarray}
\label{metrics}
\dd s^2_\pm =  -f_{\pm} \dd t^2_\pm + f_\pm^{-1}\dd r^2 + r^2 \dd \Omega_{d-1}^2,
\end{eqnarray}
where 
\begin{equation}
f_{\pm} = 1+r^2 - \dfrac{16\pi G M_{\pm}}{(d-1)V_\Omega r^{d-2}}\;, 
\end{equation}
and $V_\Omega = {\rm Vol}({\bf S}^{d-1})$. 
The shell dynamics follows from Einstein's equations, which take the form of junction conditions (cf. \cite{Israel, Poisson}). Denoting the induced metric on ${\cal W}$ as 
\begin{equation}\label{inn}
\dd s_{\cal W}^2 = - \dd \tau^2 + R(\tau)^2 \dd \Omega_{d-1}^2\, ,
\end{equation}
in terms of the shell's proper time $\tau$ and its radius $R(\tau)$, continuity of the spacetime metric across ${\cal W}$  implies the first junction condition, 
\begin{equation}
f_\pm (R) \left({\dd t_\pm \over \dd \tau}\right)^2-{1 \over f_{\pm}(R)} \left({\dd R \over \dd \tau}\right)^2 = 1\;. 
\end{equation}
 The second junction condition establishes the jump of the extrinsic curvature across ${\cal W}$ as proportional to the stress-energy on the shell's world-volume. For a thin shell of dust we have
\be\label{emt}
T_{\mu\nu} = \sigma \,u_\mu \,u_\nu \,\delta(\ell)\;,
\ee
where $u^\mu$ is the four-velocity field of the shell and $\sigma$ is the surface density. The coordinate $\ell$ measures proper distance away from ${\cal W}$ in the orthogonal spacelike direction, increasing towards the exterior region; in other words, the normal unit vector $N_{\cal W}=\partial /\partial \ell$ satisfies $N_{\cal W}^2 =1$ and $u_\mu \,N_{\cal W}^\mu =0$. For spherically infalling dust the density $\sigma (R)$  must be inversely proportional to the shell's volume, that is to say, the total rest mass 
\be\label{restm}
m= \sigma \,V_\Omega \,R^{d-1}
\ee
remains constant. 

The second junction condition specifies the jump in extrinsic curvature across ${\cal W}$, 
\be\label{second}
\sqrt{ \left({\dd R \over \dd \tau}\right)^2 + f_- (R)} - \sqrt{ \left({\dd R \over \dd \tau}\right)^2 + f_+ (R)} = {8\pi G \over d-1} \,\sigma \,R\;.
\ee 

The particular conditions of spherical symmetry and stationarity along $\mathscr{V^\pm}$ allow us to write the junction conditions in terms of the Killing vectors $\xi_\pm$, an expression that will be useful later. Using that $\xi_\mu = g_{t\mu}$ and the explicit form of the metric (\ref{metrics}) we find
\be\label{tan}
(u\cdot \xi)_\pm = -f_\pm \, {\dd t_\pm \over \dd \tau}\;.
\ee
Furthermore, since $\xi_\pm$ are orthogonal to the angular spheres, the normalization implies 
\be\label{norm}
g_{\mu\nu} \,\xi_\pm^\mu \,\xi_\pm^\nu = (\xi_\pm)^2 = -(u\cdot \xi_\pm )^2 + (N_{\cal W} \cdot \xi_\pm )^2 = -f_\pm\;,
\ee
an expression which determines $N_{\cal W} \cdot \xi_\pm $ once we know $u\cdot \xi_\pm$. Using (\ref{tan}) and (\ref{norm}) we may recast the two junction conditions as jumping rules for the Killing vectors, namely the  component  normal to ${\cal W}$ is continuous
\be\label{jumkuno}
N_{\cal W} \cdot \xi_+ \Big |_{\cal W} = N_{\cal W} \cdot \xi_- \Big |_{\cal W}\;,
\ee
whereas the component  tangential  to ${\cal W}$ jumps like the extrinsic curvature,
\be\label{jumkdos}
\left(u\cdot \xi_+ - u\cdot \xi_- \right) \Big |_{\cal W}= \sqrt{ \left({\dd R \over \dd \tau}\right)^2 + f_- (R)} - \sqrt{ \left({\dd  R \over \dd \tau}\right)^2 + f_+ (R)} = {8\pi G\over d-1} \,\sigma \,R\;.
\ee
Equivalently, we can say that both junction conditions boil down to the jump rule:
\be\label{jrule}
\left(\Delta\xi^\mu \right)_{\cal W} \equiv \left(\xi_+^\mu  - \xi_-^\mu\right)\Big |_{\cal W} = -{8\pi G \over d-1} \,\sigma \, R \,u^\mu \;.
\ee

One more presentation of the shell dynamics is obtained by extracting  from (\ref{second}) the ADM mass of the shell as a constant of motion:
\begin{equation}\label{energy}
M_{\rm shell} =M_+-M_- = m\sqrt{\left({\dd R\over \dd \tau}\right)^2 + f_-(R)} -\dfrac{4\pi G}{(d-1)V_\Omega } \dfrac{m^2}{R^{d-2}}\;.
\end{equation}
This can be interpreted as a kinetic contribution proportional to the shell's rest mass $m$, corrected by a gravitational self-energy term. In fact, the constancy of $m$ suggests a natural $(1+1)$-dimensional picture in terms of an effective particle of mass $m$, moving in the two-dimensional section of the metric obtained by simply deleting the angular directions:
\be\label{twod}
\dd s^2_{1+1} = {\bar g}_{\alpha\beta} \,\dd x^\alpha \, \dd x^\beta = -f_- (r) \dd t^2 + {\dd r^2 \over f_- (r)}\;.
\ee
 In particular, the shell energy  (\ref{energy}) can be obtained as the canonical energy from the effective   action of a free particle 
\be\label{effectivea}
S_{\rm eff} = \int \dd \lambda \,L_{\rm eff} = -m \int \dd \lambda \sqrt{{\bar g}_{\alpha\beta} {\dd x^\alpha \over \dd \lambda} {\dd  x^\beta \over \dd \lambda}}\;, 
\ee
provided we can neglect the gravitational self-energy effects.

\section{Proof of the PC duality}
\label{sec::proof}
\noindent

Our goal is to derive a PC duality relation  by direct evaluation of the left hand side of (\ref{mcd}), with ${\cal C}_{\rm shell}$ defined as in (\ref{cdef}). This will allow us to identify  the correct component of `radial momentum'. The complexity being defined through the VC prescription, we start with a preliminary discussion of extremal-volume surfaces in the relevant geometries. 

\subsection{Extremal volumes} 
\noindent

Let a codimension-one spacelike surface $\Sigma$ be defined by the embedding functions 
 $X^\mu (y^a)$, with $y^a$ coordinates along the hypersurface. The volume functional reads
\be
V\left[\Sigma \right] \, = \, \int\limits_{\Sigma}\, \dd^{d}y\, \sqrt{h} \;,\label{volumefunctional}
\ee
where $h_{ab} = \partial_aX^\mu\, \partial_bX^\nu\, g_{\mu\nu}(X)$ is the induced metric on $\Sigma$. \footnote{We use latin indices for coordinates on the hypersurface $\Sigma$ and greek indices for general coordinates in the full spacetime.}
Under a generic variation $\delta X^\mu$ the volume varies as 
\be\label{variationv}
\delta V = \int_\Sigma ({\rm e.o.m.})_\mu \,\delta X^\mu + \int_{\partial \Sigma} \dd S^a \,\partial_a X_\mu \, \delta X^\mu \;.
\ee
where 
\be\label{eom}
({\rm e.o.m.})_\mu =  -\,\dfrac{1}{\sqrt{h}}\,\partial_a\left(\sqrt{h}\,h^{ab}\,g_{\mu\nu} \,\partial_b {X^\nu} \right) \, +\, \dfrac{1}{2}\, h^{ab}\, \partial_a X^\rho\,\partial_b X^\sigma\, \partial_\mu g_{\rho\sigma} 
\ee
vanishes precisely when the hypersurface $\Sigma$ is extremal. In this case, the variation reduces to a boundary term, 
\be\label{onshellv}
\delta V \big |_{\rm extremal} = \int_{\partial \Sigma} \dd S^a\, e^\mu_a \,\delta X_\mu\;,
\ee
where we have defined the vector fields $e^\mu_a = \partial_a X^\mu$ tangent to $\Sigma$.

\begin{figure}[h]
	\centering
	\includegraphics[width = .65\textwidth]{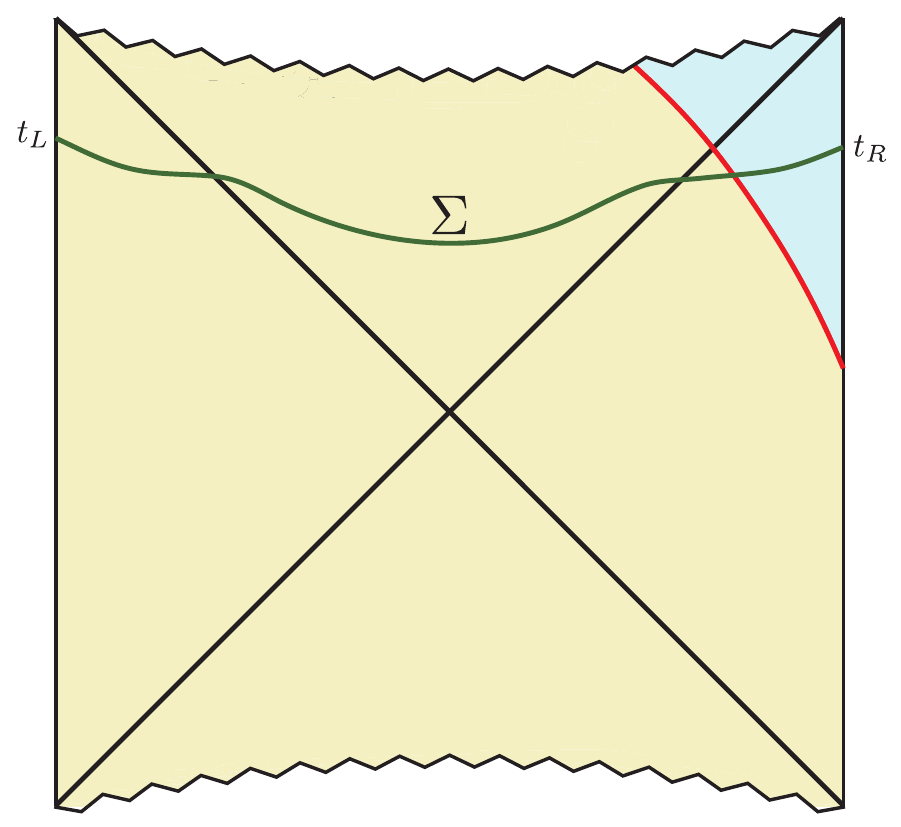}
	\caption{Extremal codimension-one surface $\Sigma$ of interest. Its boundary $\partial \Sigma$ consists of two spheres at infinity, located at times $t_L = t_R = t$.}
	\label{maximalslices}
\end{figure}

For the geometry of interest here, $
\Sigma$ is a cylindrical manifold of topology ${\bf R} \times {\bf S}^{d-1}$, the boundary having two disconnected components consisting of spheres at the left and right spatial infinities. We shall use the same future-directed time variables on both boundaries and take a left-right symmetric time configuration  $t_L = t_R =t$, so that we can write the following boundary conditions at the regularization surfaces $r=r_\Lambda$: 
\be\label{bcon}
\delta X^\mu_\pm \Big |_{r=r_\Lambda}   = \pm \delta t\; \xi_\pm^\mu \Big |_{r=r_\Lambda}\;,
\ee
where the $\pm$ signs account for the fact that the left-side Killing vector $\xi_-$ is past-directed at large radii. Spherical symmetry allows us to parametrize the induced metric on extremal surfaces in the form
\be\label{induced}
\dd s^2_\Sigma = h_{ab} \,\dd y^a \,\dd y^b = \dd y^2 + g(y) \,\dd \Omega^2_{d-1}\;,
\ee
where $y$ is a radial coordinate running over the real line, with $y=\pm\infty$ corresponding  respectively to the left and right boundaries of $\Sigma$. 
In these coordinates, we can picture $e_y^\mu = \partial_y X^\mu$ as a unit-normalized, radial, spherically symmetric, right-pointing vector field. Denoting the spheres at  infinity by ${\cal S}_{\pm\infty}$ we can rewrite the volume variation of extremal surfaces (\ref{onshellv}) as
\be\label{moreonshell}
\delta V \big |_{\rm extremal} = \delta t\, \left[\int_{{\cal S}_{+\infty}} e_y^\mu (\xi_+)_\mu + \int_{{\cal S}_{-\infty}} e_y^\mu \,(\xi_-)_\mu \right]\;,
\ee
where we have absorbed the sign assignments in (\ref{bcon}) into a reversal of orientation for the left-boundary integral. Namely, both integrals in (\ref{moreonshell}) are now written as scalar integrals over the boundary spheres. 

This expression for  the volume dependence with asymptotic time is useful because the featured integrals turn out to be Noether charges. If we view the volume functional (\ref{volumefunctional}) as an action on a collection of fields $X^\mu$ defined over $\Sigma$, the isometries of the $\mathscr{V^\pm}$ portions  are interpreted as `internal symmetries' of the this field theory, with their corresponding Noether currents. The time-translation symmetries associated to
$\xi_\pm$ induce Noether currents of the form \footnote{In order to prove conservation, we just use $\xi_\mu = g_{t\mu}$ and evaluate the equation of motion from (\ref{eom}).}
\be\label{noether}
J_a = e_a^\mu \,\xi_\mu\;,\qquad \nabla_a J^a =0\;.
\ee
In particular, the integral of the radial component $J_y$ over any fixed-$y$ section ${\cal S}_y$ is a Noether charge which is conserved under transport in the $y$ direction:
\be\label{nocha}
\Pi[{\cal S}_y] = \int_{{\cal S}_y} e_y^\mu\; \xi_\mu \;, \qquad \partial_y \Pi[{\cal S}_y] =0\;.
\ee

\subsection{Identification of the PC component}
\label{sec::identification}
\noindent

We have  now the machinery in place to evaluate (\ref{cdef}). The formula (\ref{moreonshell}) implies 
\be\label{pis}
{\dd V \over \dd t} = \Pi_+ + \Pi_- \;,
\ee
in terms of the Noether charges  $\Pi_\pm \equiv \Pi[{\cal S}_{\pm \infty}]$ on right and left boundaries (a similar result was derived in \cite{MyersVaidyaI, MyersVaidyaII} for null shells). The normalization of the operator complexity requires the subtraction of the same expression, evaluated on the Noether charges $\Pi_\pm^{(0)}$  of the eternal black hole geometry without infalling shell, namely
\be\label{cdefn}
{\dot{\cal C}}[{\cal O}_{\rm shell}]  = {d-1 \over 8\pi G} \left[ \Pi_+ - \Pi_+^{(0)} + \Pi_- - \Pi_-^{(0)} \right]\;,
\ee
where the dot here denotes derivative with respect to asymptotic time. 

Left-right symmetry of the eternal black hole geometry implies $\Pi_+^{(0)} = \Pi_-^{(0)}$, whereas we can also set $\Pi_- \approx \Pi_-^{(0)}$ at the left regularization boundary because, for shells that enter the geometry at very late times,  their  worldvolume ${\cal W}$ remains very far from the left boundary. Hence, near the left regularized boundary,  the extremal surface $\Sigma$ is very well approximated by that of the eternal black hole. As we remove the regularization, in the limit $r_\Lambda \rightarrow \infty$, we must actually obtain $\Pi_- =  \Pi_-^{(0)}$. This allows us to  remove all explicit reference to the eternal black hole geometry and write
\be\label{unamas}
 {\dot{\cal C}}[{\cal O}_{\rm shell}] = {d-1 \over 8\pi G} \left[ \Pi_+ - \Pi_- \right]\;.
\ee
Furthermore, the conservation of Noether charges in either  $\mathscr{V^+}$  or  $\mathscr{V^-}$  allows us to bring the Noether charges to both sides of the shell's worldvolume:
\be\label{fins}
 {\dot{\cal C}}[{\cal O}_{\rm shell}]= {d-1 \over 8\pi G}(\Delta \Pi )_{\cal W}=  {d-1 \over 8\pi G}  \int_{{\cal S}_{\cal W}} e_y^\mu \,(\Delta \xi_\mu)_{\cal W}\;,
\ee
where $(\Delta \xi^\mu)_{\cal W} = (\xi_+^\mu - \xi_-^\mu)\big |_{\cal W} $ is the jump of the Killing vectors across ${\cal W}$ and ${\cal S}_{\cal W}$ is the sphere at the intersection $\Sigma \cap {\cal W}$. Using now the junction conditions in the form (\ref{jrule}), we find 
\be\label{fine}
{\dot{\cal C}}[{\cal O}_{\rm shell}]= -\int_{{\cal S}_{\cal W}}  \sigma\,R\, e_y^\mu \,u_\mu \;.
\ee
We can now elaborate (\ref{fine}) in various ways in order to flesh out the PC-duality interpretation. First, we define a `complexity field' over $\Sigma$ as a rescaling of the $e_y^\mu$ field:
\be\label{compf}
{\cal C}_{\Sigma}^\mu \equiv -r\,e_y^\mu\;.
\ee
Second, we define a density of proper momentum along the shell's worldvolume 
\be\label{properd}
{\cal P}^\mu \equiv \sigma \,u^\mu\;.
\ee
With these definitions we can rewrite (\ref{fins}) as 
\be\label{pdun}
{\dot{\cal C}}[{\cal O}_{\rm shell}]= P_{\cal C} = \int_{{\cal S}_{\cal W}}  {\cal P}_\mu \,{\cal C}_{\Sigma}^\mu \;,
\ee
a relation which identifies the precise component of momentum which is dual to complexity growth, namely the projection of the proper momentum along the direction of the complexity vector field ${\cal C}_{\Sigma}^\mu$. It is a particular radial component with inward orientation and appropriate normalization. 

\begin{figure}[h]
	\centering
	\includegraphics[width = 0.46\textwidth]{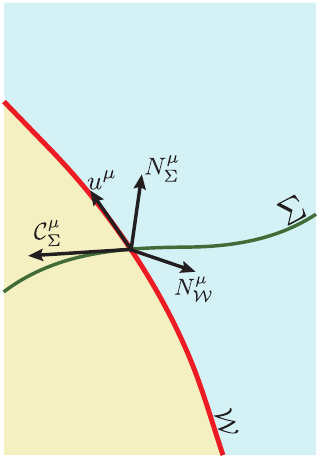}
	\caption{Configuration of relevant vectors at the intersection sphere $\mathcal{S}_{\mathcal{W}} = \Sigma \cap {\cal W}$.}
	\label{diagram}
\end{figure}

\vspace{1mm}
A second presentation of this result has the virtue of hiding some of the peculiarities of the concrete system we have considered so far. In fact, no explicit geometrical information about the shell's worldvolume ${\cal W}$ is needed in order to express the PC duality relation. To see this, let us consider the expression
\be\label{emflux}
-\int_\Sigma N_\Sigma^\mu \;T_{\mu\nu} \;{\cal C}_{\Sigma}^\nu\;,
\ee
where $N_\Sigma$ is the unit timelike normal to $\Sigma$. It measures the flux through $\Sigma$ of a suitably normalized momentum component along $\Sigma$. Upon explicit evaluation for the spherical shell, using (\ref{emt}), we find
\be\label{eval}
-\int \dd y \int_{{\cal S}_y} \, \sigma\, (N_\Sigma \cdot u) \,({\cal C}_{\Sigma} \cdot u) \,\delta(\ell) \;. 
\ee
Furthermore, $\delta (\ell) = \delta(y-y_{\cal W}) \,|d\ell /dy |^{-1}$, where $y_{\cal W}$ is the value of the $y$ coordinate at the shell's intersection. From the definition of the ${\cal W}$-normal we have $ d\ell /dy =  \partial_y X^\mu\,\partial_\mu \ell \, = e_y\cdot N_{\cal W} $, which allows us to collapse the integral to the intersection sphere ${\cal S}_{\cal W}$:
\be\label{evalmore}
\int_{{\cal S}_{\cal W}} \sigma\, R\,{(N_\Sigma \cdot u)\, (e_y \cdot u) \over (e_y \cdot N_{\cal W})}\;,
\ee
where we have used (\ref{compf}). To further reduce this integral we notice that $N_\Sigma$ and $e_y$ are orthogonal and unit normalized, as well as the pair $u$ and $N_{\cal W}$, so that we have $N_\Sigma \cdot u= - N_{\cal W} \cdot e_y$, where the minus sign accounts for the timelike character of both $N_\Sigma^\mu$ and $u^\mu$. This simplifies (\ref{evalmore}) and recovers (\ref{fine}). Hence, we have established the more intrinsic form of the PC relation:
\be\label{guay}
 {\dot{\cal C}}[{\cal O}_{\rm shell}]= P_{\cal C} = -\int_\Sigma N_\Sigma^\mu \;T_{\mu\nu} \;{\cal C}_{\Sigma}^\nu\;.
\ee

In this version, all explicit reference to the details of the bulk state gets reduced to its stress-energy tensor. The vector fields $N_{\Sigma}$ and ${\cal C}_{\Sigma}$ are defined in terms of the extremal surface, whose detailed geometry is also determined by $T_{\mu\nu}$ through the back reaction on the geometry. Indeed, the form of (\ref{guay}) should remain valid for spherical shells with any internal equation of state, including those corresponding to branes which change the AdS radius of curvature across ${\cal W}$. Furthermore, the role of the Noether charges in the derivation of \eqref{pdun} and \eqref{guay} makes it clear that it applies as well to spherical thin shells collapsing in vacuum AdS and forming a one-sided black hole. 

More generally, we expect that any spherical matter distribution can be approximated by a limit of many concentric thin shells, so that
(\ref{guay}) should remain valid for {\it any} matter bulk distribution  with spherical symmetry. It would be interesting to have a direct derivation of this fact,
which could shed light on whether (\ref{guay}) remains true without spherical symmetry. The generalization to one-sided collapse of thin shells with  arbitrary equations of state, but still maintaining spherical symmetry, is explained in Appendix \ref{sec::appendixB}. A first step towards lifting the spherical symmetry requirement is presented in Appendix \ref{sec::appendixC}, which considers a formal collapse of a rotating shell in AdS$_3$.

\section{Late time limit and the black hole interior}
\noindent

One chief motivation behind this work is the elucidation of the very late time regime of operator complexity growth in the light of the PC duality. Any definition of operator complexity with the structure of equation (\ref{opsub}) will assign a linear growth at late times. In particular, given that state complexities are expected to grow proportionally to  $E_\Psi \,t$, where $E_\Psi$ is a characteristic energy of the state, the subtracted definition for operator complexity gives a slope proportional to $E_{\cal O} \,t$, where $E_{\cal O}$ is the  additional energy injected by the operator ${\cal O}$. Translated to our gravitational set up, we expect a late time behavior
\be\label{latet}
{\dot{\cal C}}[{\cal O}_{\rm shell}]\Big |_{\rm late} \approx M_+ - M_- = M_{\rm shell}\;.
\ee
We would like to check that our PC relation satisfies  this expected asymptotic behavior. A simple check can be performed in the limit of very large AdS black holes. This coincides with the situation where the infalling shells have small gravitational self-energy at all times that are relevant for the calculation.

\begin{figure}[h]
	\centering
	\includegraphics[width = .9\textwidth]{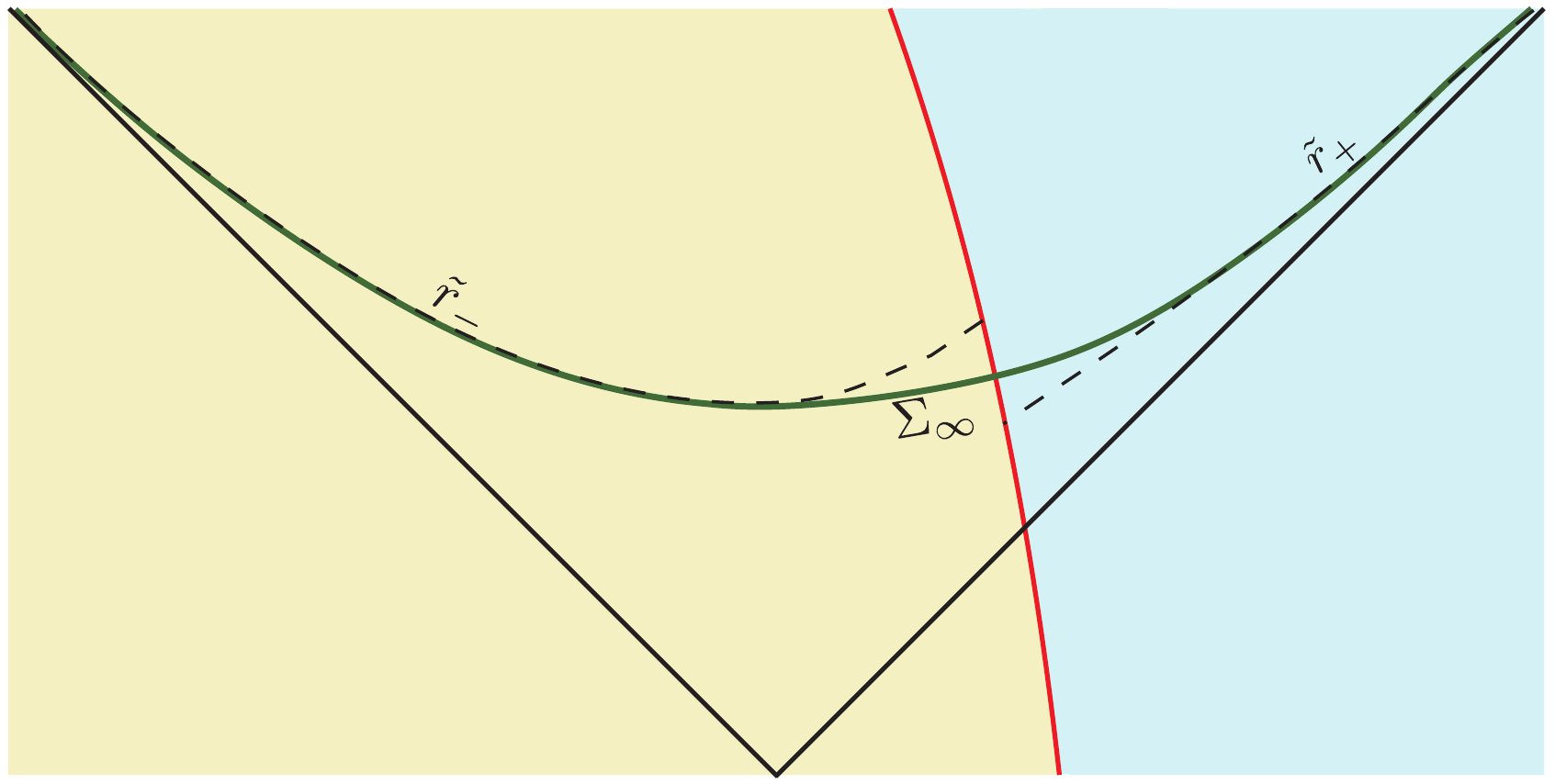}
	\caption{The saturation slice $\Sigma_{\infty}$ interpolates between the extremal surface barrier inside $\tilde{r}_{-}$ and outside $\tilde{r}_{+}$.}
	\label{lastslice}
\end{figure}

The key point is to notice that, at late times, the extremal surfaces $\Sigma_t$ accumulate in the interior of the black hole, exponentially converging \footnote{See Appendix \ref{sec:appendix} for an quantitative discussion of this phenomenon.} to a limiting surface $\Sigma_\infty$ (cf. \cite{SusskindNotEnough, WallEngelhardt}). For a shell that enters the black hole very late, this surface interpolates between
the limiting surfaces $(\Sigma_\infty)_\pm$ associated to the early and late black holes of mass $M_\pm$ (cf. Figure \ref{lastslice}). In terms of the interior Schwarzschild radial coordinates, let ${\tilde r}_\pm$ denote the saturation radii, defined by the local extremization of the `volume Lagrangian'
$
r^{d-1} \sqrt{|f(r)|} 
$. By explicit calculation we find, in the limit of very large AdS black holes 
\be\label{largea}
{\tilde r}^{\;d} \approx {8\pi G M\over (d-1) V_\Omega} \;.
\ee

We can now make use of the `movability' of the Noether charges $\Pi_\pm$ to evaluate then away from ${\cal W}$, but still inside the black hole interior,
in a region where $\Sigma_t$ is well-approximated by a constant-$r$ surface. Let us denote the angular spheres at such points by ${\widetilde {\cal S}}_\pm$.  Then, equation (\ref{unamas}) can be rewritten as 
\be\label{otramas}
 {\dot{\cal C}}[{\cal O}_{\rm shell}]\Big |_{\rm late} \approx {d-1 \over 8\pi G} \,\left( \Pi\left[{\widetilde {\cal S}}_+ \right] - \Pi\left[{\widetilde {\cal S}}_-\right] \right)\;.
 \ee
 In computing the Noether charges, we notice that $\xi_\pm = \partial /\partial t_\pm$ are approximately tangent to $\Sigma_t$ in the saturation region. Hence, we can write $e_y^\mu \approx \xi^\mu /\sqrt{\xi^2}$ and the Noether integrals are simply
 \be\label{noetin}
 \Pi[{\widetilde {\cal S}}_\pm] \approx \int_{{\widetilde {\cal S}}_\pm} \sqrt{ \xi^2} = V_\Omega \,{\tilde r}_\pm^{\,d-1} \sqrt{ |f({\tilde r)}_\pm|} \approx  V_\Omega \, {\tilde r_\pm}^{\;d} \approx {8\pi G M_\pm \over d-1}\;.
 \ee
 In the last equality we have made use of (\ref{largea}) and the approximation of a large AdS black hole.  Therefore, upon subtraction we  conclude the proof of (\ref{latet}). 
 
 An important observation regarding this result is the fact that the vector fields ${\cal C}^\mu$ and $e_y^\mu$ do differ significantly in the interior saturation region, because the rescaling factor ${\tilde r}$ is non trivial there, and yet this rescaling is crucial to obtain the expected long-time asymptotics. Therefore,
 the peculiar normalization (\ref{compf})  of the momentum component along $\Sigma$ is necessary for 
 the consistency of the results. 
 
 We can obtain further insight into the rationale behind the linear complexity growth by passing to the effective particle description. Again neglecting self-energy corrections, we can envision the dynamics of the shell as that of a probe particle of mass $m$ falling through the $(1+1)$-dimensional metric (\ref{twod}). The PC duality relation admits the two-dimensional representation:
 \be\label{twopc}
  {\dot{\cal C}}[{\cal O}_{\rm shell}] = P_{\cal C} = P_\alpha \,{\cal C}^\alpha\;,
  \ee
  where $P^\alpha = m \,u^\alpha$, with  $\alpha$  a two-dimensional index. Picking for example the standard $(r,t)$ coordinates, we have  
 \be\label{twoc}
 P_{\cal C} = -r\left({\partial t \over \partial y} \,P_t + {\partial r \over \partial y} P_r \right)
 \;.
 \ee
 Let us introduce an adapted coordinate for the radial `complexity field' ${\cal C}^\alpha =-r e^\alpha_y$, namely we define a rescaled radial coordinate $\chi$ such that
 \be\label{chi}
 {\cal C}^\alpha = \left({\partial \over \partial \chi}\right)^\alpha = -r \,e^\alpha_y = -r \,\left({\partial \over \partial y}\right)^\alpha\;, 
 \ee
 or, equivalently 
 \be\label{defchi}
 {\partial \over \partial \chi} = - r\,{\partial \over \partial y}\;.
 \ee
 Using the so-defined $\chi$ coordinate, we can simplify (\ref{twoc}) so that
 \be\label{chip}
 P_{\cal C} = P_t \,{\partial t \over \partial \chi} + P_r \,{\partial r \over \partial \chi} = P_\chi\;.
 \ee
 To the extent that we are only interested in describing the particle motion to the past of the saturation surface $\Sigma_\infty$,  we may use a time slicing given by the extremal surfaces $\Sigma_t$  themselves, and coordinate the spacetime
 in terms of $(t, \chi, \Omega)$. In this frame, the complexity momentum coincides with the $\chi$-canonical momentum, provided we stay within  the probe approximation:
 \be\label{cancan}
 P_{\cal C} = P_\chi = {\partial L_{\rm eff} \over \partial {\dot \chi}}\;.
 \ee
 This brings our general formalism into contact with the discussion of canonical Rindler momentum in the introduction.  However, the present treatment is capable of describing the late-time behavior of the complexity. In particular, the use of a time slicing adapted to the extremal surfaces leads to the phenomenon of {\it saturation} in the black-hole interior. This freezes the value of the momentum at a constant value for asymptotically large values of $t$, thereby explaining why a linear growth of  complexity can be compatible with a PC-type formula (\ref{mcd}). 
 
\begin{figure}[H]
	\centering
	\includegraphics[width = .75\textwidth]{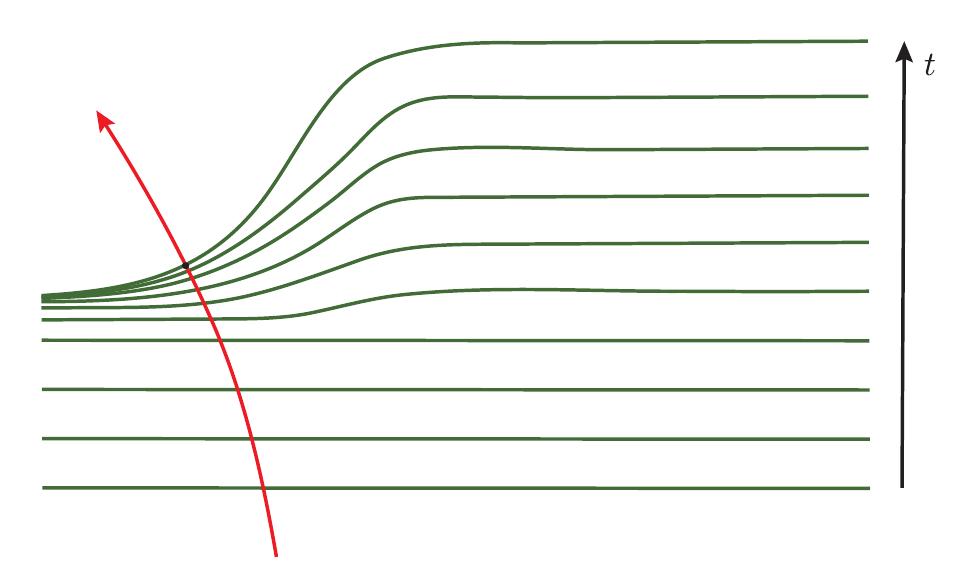}
	\caption{The  late-time saturation of the time slicing in the interior of the black hole results in a frozen momentum component, as required for any PC formula which should apply in a regime of linear complexity growth.}
	\label{saturation}
\end{figure}

\section{Conclusions and Outlook}
\noindent

In this paper we have presented a bulk derivation of a particular version of the momentum/complexity (PC) duality which applies to arbitrary times. By examining the VC complexity of thin spherical shells impinging on double-sided AdS black holes, we can explicitly identify the relevant momentum component. 

The key to the construction is to measure this momentum with respect to a bulk time foliation by the same maximal surfaces that one uses to compute VC complexity. This allows us to express the PC relation in the form 
\be\label{spheg}
{\dot {\cal C}}[{\cal O}_{\rm spherical}] = - \int_{\Sigma} N_\Sigma^\mu \,T_{\mu\nu} \,{\cal C}_\Sigma^\nu \;, 
\ee
so that the dynamical properties of the shells only enter through the energy momentum tensor $T_{\mu\nu}$. All other objects appearing in this formula are defined in terms of the chosen time foliation by maximal hypersurfaces. Although the formal relation  between VC of shells and appropriate canonical momenta has appeared before in various works (cf. for example \cite{SusskindZhao, MyersTimeDependence, MyersVaidyaI, MyersVaidyaII}),  the structure of (\ref{spheg}) suggests that it should generalize much beyond the context of thin shells, since all data entering the right hand side of (\ref{spheg})  actually  make sense for arbitrary bulk states. 

It would be interesting to find a general {\it ab initio} derivation of this relation which does not go through the thin-shell detour (cf. \cite{toappear}) The arguments presented in this paper  do apply to
collapsing thin shells in the AdS vacuum \footnote{See Appendix \ref{sec::appendixB} for a more detailed derivation of the one-sided PC formula.}. In this case, gravitational self-energy cannot be neglected at the saturation surface in the resulting one-sided black hole, so that one expects the probe approximation to be less efficient in the effective particle picture. 

It  would be interesting to check the complexity slope (\ref{latet}) by direct evaluation of $P_{\cal C}$. This requires detailed control of the precise location of the intersection sphere ${\cal S}_{\cal W}$ in the black-hole interior.  It is also interesting to check whether a transient exists for early times which shows a measurable Lyapunov exponent. This is a nontrivial fact, given that our time foliation is quite different from a near-horizon Rindler system.  On the other hand, the occurrence  of such a  transient  with exponential growth is independent of our particular PC correspondence, which is only a rewriting of the standard VC complexity formula.  In particular, such chaotic transients were numerically identified    in \cite{MyersVaidyaII, Thorlaciusdilaton} in VC computations relevant to situations which are similar, although not identical, to the set up studied in this paper.

One outstanding   question raised by our results is the true generality of a formula like (\ref{spheg}). In particular, its validity for non-spherical situations and the elucidation of the deeper geometrical meaning of the `complexity vector field' ${\cal C}^\mu_\Sigma$. Appendix \ref{sec::appendixC} presents a formal solution for a collapsing shell in AdS$_3$ with rotation, where it is found that the complexity rate still satisfies (\ref{spheg}) with the {\it same} complexity field in this less symmetrical situation. 

Even more generally, it would be interesting to explore versions of the PC duality in spacetimes without any Killing vectors. In particular, gravitational radiation corrections to (\ref{spheg}) should be accessible by perturbative methods around the solutions discussed in this paper. Finally,  it is natural to expect that (\ref{spheg})  is related to the  so-called `first law of holographic complexity' \cite{Firstlaw1, Firstlaw2}  in ways that should be elucidated more precisely.

\section{Acknowledgments}
\label{ackn}
\noindent

We would like to thank  C. Gomez, J. Magan, E. Rabinovici, R. Shir and  R. Sinha  for discussions on various aspects of computational complexity. This work is partially supported by the Spanish Research Agency (Agencia Estatal de Investigaci\'on) through the grants IFT Centro de Excelencia Severo Ochoa SEV-2016-0597,  FPA2015-65480-P and PGC2018-095976-B-C21. The work of J.M.G. is funded by 
Fundaci\'on La Caixa under ``La Caixa-Severo Ochoa'' international predoctoral grant. The work of M.S. is funded by the FPU Grant FPU16/00639.

\cleardoublepage

\appendix

\section{Late time accumulation of maximal slices}
\label{sec:appendix}
\noindent

In this appendix, we show proof of the exponentially fast accumulation of maximal slices in the black hole interior. For that matter, we will work within the benchmark case of an eternal black hole, whose metric is given in Eddington-Finkelstein coordinates by
\be
\dd s^2 \, = \, -f(r)\, \dd u^2 \, +\,2\, \dd u\, \dd r + r^2 \dd \Omega_{d-1}^2 \;.
\ee

By spherical symmetry, the maximal surface can be written as a direct product $\Sigma= \gamma \times {\bf S}^{d-1}$, with $\gamma$ a curve in the $u-r$ plane. Exploiting this symmetry we can reduce thus the problem of volume extremalization to that of a spacelike geodesic in the effective two-dimensional spacetime
\begin{equation}
\dd s_\gamma^2 = \, r^{2(d-1)}(-f(r)\, \dd u^2 \, +\,2\, \dd u\, \dd r ) \;,
\end{equation}
so that the effective volume functional is given by
\be
 V[\Sigma] V_\Omega^{-1}\, =V[\gamma]= \, \int \dd \lambda \; r^{d-1} \sqrt{-f(r)\,\dot{u}^2 + 2\,\dot{u} \,\dot{r} }\;,\label{volumetwosided}
\ee
where $\lambda$ is an arbitrary spacelike parameter and the dot stands for $\dd / \dd \lambda$. The Lagrangian in \eqref{volumetwosided} enjoys a conserved charge associated to the static Killing
\be
\label{Pidef}
\Pi \,= \dfrac{ \partial \mathcal{L}_\gamma}{\partial \dot u} = \, r^{d-1}\, \dfrac{-f(r)+\dot r}{\sqrt{-f(r)+ 2\,\dot r}}\;,
\ee
where  $\Pi$ is guaranteed to be positive by the spacelike character of the geodesic $\dd s^2_\gamma >0$ and we have taken the convinient gauge choice $\lambda = u$. Feeding the conserved charge into the equations of motion for $r(u)$ we get
\be
\dot r \,=\, f(r) \,+\, \dfrac{\Pi^2}{r^{2(d-1)}} \,+\, \dfrac{\Pi}{r^{d-1}}\, \sqrt{\dfrac{\Pi^2}{r^{2(d-1)}} +f(r)}\;. \label{motiontwosided}
\ee

Upon the imposition of reflection symmetry in our setup ($t_L=t_R=t$), the boundary conditions can be recasted to be $\dot{r}(u_i)=0$ and  $r(u_\infty)=r_\infty$ for $u_i = r_*(r_i)\; , u_\infty=t$ the values of the parameter at the symmetric turning point and boundary respectively.  In terms of the turning point radius $r_i$ we can get a simple expression for $\Pi$
\be
\Pi\,  =\, r_i^{(d-1)}\, \sqrt{-f(r_i)} \;. \label{definitionsigmatwosided}
\ee

An implicit relation between $t$ and $r_i$ can be obtained integrating \eqref{motiontwosided} 
\be
\int^{u_{\infty}}_{u_i} du \; = \; \int^{r_\infty}_{r_i} dr \, \dfrac{r^{2(d-1)}}{g^{1/2}(r)\, \left(\Pi+ g^{1/2}(r)\right)} \;.\label{int1}
\ee
where we have defined the function
\be
g(r) = \,r^{2(d-1)}\, f(r)\,-\,r_i^{2(d-1)}\, f(r_i)\;,
\ee
which vanishes at the minimal radius $r_i$. Breaking up the radial integral into an inner an outer piece and substituting the boundary conditions, we can obtain an expression for the boundary time
\be
t\; = \; \int^{r_h}_{r_i} dr \, \dfrac{r^{2(d-1)}}{g^{1/2}(r)\, \left(\Pi+ g^{1/2}(r)\right)} + h(r_h,r_i, r_\infty) \;.\label{intt}
\ee
where $h(r_h,r_i, r_\infty)$ is a finite function for all values of its parameters. As we see from  the structure of the zeros of $g(r)$, the integral above contains a pole at $r=r_i$. In order to approximate the integral \eqref{intt} we may expand $g(r)$ to second order around $r_i$
\be
g(r) = \, \alpha (\tilde{r}_i-r_i) (r-r_i) \,+\, \frac{\alpha}{2} (r-r_i)+ \, ...\;. \label{expansionsecondorder}
\ee
where $\alpha$ is a positive constant depending on the parameters of the black hole and $\tilde{r}_i$ is the asymptotic limiting surface. The necessity to go up to second order in the expansion is revealed by the vanishing of the linear term in the late time limit corresponding to $r_i \rightarrow \tilde{r}_i$. Feeding \eqref{expansionsecondorder} into \eqref{intt} and expanding the rest of the integral to zero order we get
\be
t\; \approx \; \dfrac{r_i^{2(d-1)}}{\Pi} \int^{r_h}_{r_i} dr \, \left[\alpha (\tilde{r}_i-r_i) (r-r_i) \,+\, \frac{\alpha}{2} (r-r_i)\right]^{-1/2}  \;+\;\text{finite} \;.
\ee
which can be solved exactly
\be
t \; \approx \; -\dfrac{r_i^{2(d-1)}}{\Pi (\alpha /2)^{1/2}}\,\log (r_i-\tilde{r}_i) \;+\;\text{finite} \;.
\ee

Inverting this expression we get the desired result, i.e. the exponentially fast saturation of maximal slices in the black hole interior
\be
r_i-\tilde{r}_i \; \approx \; b\, e^{-t/a}\ \;, \label{exponentialaccumulation}
\ee 
where $a$ and $b$ approach constant values in the late time limit.

\cleardoublepage

\section{One-sided PC duality}
\label{sec::appendixB}
\noindent

In this appendix, we extend the regime of validity of the PC duality \eqref{guay} to situations in which there is a spherically symmetric thin shell living in the AdS vacuum. We  introduce a slightly more general formalism to manifestly show that the same PC formula holds for any spherical thin shell irrespectively of its internal equation of state. 

We start from a single holographic CFT on ${\bf S}^{d-1}$ and take the CFT vacuum as the reference state to define the operator complexity \eqref{opsub}. Using the VC prescription, the bulk definition is 
\be\label{complexdefonesided}
{\cal C}\left[{\cal O}_{\rm shell}\right] = {d-1 \over 8\pi\, G } \left[ {\rm Vol} (\Sigma_{\rm AdS+shell}) - {\rm Vol}(\Sigma_{\rm AdS})\right]\;,
\ee
where $\Sigma$ is the extremal hypersurface of interest, defined in empty AdS with and without the shell respectively. A peculiarity of this choice of reference state is that its complexity is constant in time, and this makes the rate of \eqref{complexdefonesided} to depend only on the extremal hypersurface on the spacetime with the shell. This extremal volume hypersurface $\Sigma$ will be topologically a ball anchored to the asymptotic sphere ${\cal S}_{\infty}$ at boundary time $t$. A generic infinitesimal deformation of its embedding function $\delta X^\mu = \delta\varepsilon\, N_\Sigma^\mu \, + \, \delta\kappa^a \, e^\mu_a$ will produce the volume variation
\be
\delta V[\Sigma]\big |_{\rm extremal}  \, = \, \int_{\Sigma}\,  \nabla_{a}\delta\kappa^{a}\, = \int_{{\cal S}_{\infty}}\,dS_a\,\delta\kappa^a\;, \label{var1}
\ee
as in \eqref{onshellv}, which in this case follows from the tracelessness of the extrinsic curvature of $\Sigma$. In particular, for time translations of the boundary sphere, we need to take the tangent deformation to asymptotically become $(\delta \kappa_a)|_{{\cal S}_{\infty}}=  (\partial_t \cdot  e_a) \,\delta t \label{asymptotic}$. From \eqref{var1}, the rate of extremal volume then reads 
\be
\dfrac{\text{d} V}{\text{d}t}\, = \, \int_{\Sigma}\,  \nabla_{a}\,\rho^{a}\, \equiv \, \Pi\;,\label{evolution}\,
\ee
for $\rho^a$ any tangent vector that asymptotically approaches $\partial_t \cdot  e^a$.

For spherically symmetric thin shell configurations, there will be two timelike Killing vectors $\xi^\mu_\pm$ individually defined on each of the regions of spacetime $\mathscr{V^\pm}$ glued by the worldvolume $\mathcal{W}$. Taking $\ell$ as a normal coordinate to $\mathcal{W}$, we can define the Killing vector field globally as $\xi^\mu =  \xi_-^\mu \,\Theta(-\ell)\, +\, \xi_+^\mu \,\Theta(\ell)$,
where $\Theta$ is the step function. The Killing condition is then broken due to a possible discontinuity across $\mathcal{W}$
\be
\nabla_{(\mu}\,\xi_{\nu)}\, = \, \left(N_\mathcal{W}\right)_{(\mu}\,\left(\Delta \xi\right)_{\nu)}\,\delta(\ell)\;, \label{Killingbroken}
\ee
where we used that $\partial_\mu \Theta(\ell) = \delta(\ell)\left(N_\mathcal{W}\right)_\mu\,$, for $N_\mathcal{W}^\mu$ the $\mathcal{W}$-normal. The global piecewise Killing $\xi^\mu$ asymptotically becomes the time translation generator $\partial_t^\mu$, and therefore it is possible to choose its projection to the extremal hypersurface $\xi^a = \xi \cdot  e^a$ to play the role of the tangent vector in \eqref{evolution}. The projection of \eqref{Killingbroken} into $\Sigma$ reads
\be
\nabla_{(a}\,\xi_{b)} \, =\, \delta(\ell)\, \left(N_\mathcal{W}\cdot e_{(a}\right)\,\left(\Delta \xi\cdot e_{b)}\right)\, + \, (\xi\cdot N_\Sigma)\,K_{ab}\,\;,
\ee
where $K_{ab}$ the extrinsic curvature of $\Sigma$. This second term breaks the Killing condition as a consequence of the original Killing $\xi^\mu$ failing to be tangent to $\Sigma$. Nevertheless, for the extremal hypersurface $\Sigma$ the trace of this term vanishes, which makes this tangent vector to be conserved away from $\mathcal{W}$ 
\be
\nabla_{a}\,\rho^{a} \, =\, \delta(\ell)\, \left(N_\mathcal{W}\cdot e_{a}\right)\,h^{ab}\,\left(\Delta \xi\cdot e_{b}\right)\;.
\ee
This tangent vector precisely agrees with the Noether current \eqref{noether} arising from the internal time-translation symmetry of the volume functional.

In this framework, we thus find that the rate of the operator complexity is proportional to a localized quantity on $\mathcal{W}$
\be
\dot{\mathcal{C}}\left[{\mathcal{O}}_{\text{shell}}\right]\, \, =\,\dfrac{d-1}{8\pi G}\,\int_{\Sigma}\,  \delta(\ell)\, \left(N_\mathcal{W}\cdot e_{a}\right)\,h^{ab}\,\left(\Delta \xi\cdot e_{b}\right)\;,\label{stationarypi}
\ee
namely the discontinuity of the stationary Killing vector field. 

To evaluate the discontinuity of the Killing vector across $\mathcal{W}$, let us focus on the codimension two sphere of intersection $\mathcal{S}_\mathcal{W} =  \Sigma \cap \mathcal{W}$. We define the spacelike tangent to $\Sigma$ which is orthogonal to $\mathcal{S}_\mathcal{W}$ and unit norm, denoted by $e_y^\mu$. Similarly, we define the timelike tangent to $\mathcal{W}$, denoted $u^\mu$, as the one orthogonal to $\Sigma \cap \mathcal{W}$ and unit norm. From spherical symmetry $\xi^\mu_\pm|_{\mathcal{W}}$ will be orthogonal to the spheres, and an identical argument to the one provided in section \ref{sec:collapse} determines that the only discontinuity will be tangent to $\mathcal{W}$ and with value
\be
\left(\Delta \xi^{\mu}\right)_{\mathcal{W}}\, = \, -\dfrac{8\pi G}{d-1}\left(S_{\rho\sigma}\,u^\rho\, u^\sigma\,R\right) \, u^\mu\;,
\ee
where $S_{\mu\nu}$ is the induced energy-momentum on $\mathcal{W}$, and $R$ is the radius of $\mathcal{S}_\mathcal{W}$. Substituting in \eqref{stationarypi} and noting that $N_\mathcal{W}\cdot e_{y}$ can be written as $- N_\Sigma \cdot u$ from the argument given in \ref{sec::identification}, we get
\be
\dot{\mathcal{C}}\left[{\mathcal{O}}_{\text{shell}}\right]\, \, =\,\,\int_{\Sigma}\,  \left(T_{\mu\nu} u^\mu u^\nu\right)\,r\, \left(N_{\Sigma}\cdot u\right)\left(u\cdot e_{y}\right)\;.\label{stationarypi2}
\ee

The one-sided version of the PC duality then follows from the decomposition $u^\mu u^\nu  =  -g^{\mu\nu}  +N^{\mu}_\mathcal{W}N^\nu_\mathcal{W}+ g^{\mu\nu}_{\mathcal{S}_{\mathcal{W}}}$, where the last term is the induced metric on $\mathcal{S}_{\mathcal{W}}$, and from the thin-shell condition $T_{\mu\nu}\, N^\nu_{\mathcal{W}} = 0$. Upon the definition of the `complexity field' ${\cal C}_{\Sigma}^\mu\, = \, -r\,e_y^\mu$, we arrive at the desired formula
\be\label{bdiez}
{\dot{\cal C}}[{\cal O}_{\rm shell}]\, \, =\,-\int_{\Sigma}\,  N_\Sigma^\mu \;T_{\mu\nu} \;{\cal C}_{\Sigma}^\nu\;.
\ee

This derivation of the PC duality certainly clarifies that the PC formula applies to any spherically symmetric thin shell in AdS, including branes that separate  AdS patches of different curvature radius.

\section{Rotating thin shell in AdS$_3$}
\label{sec::appendixC}
\noindent

In this Appendix we use the language developed in Appendix \ref{sec::appendixB} to begin exploring less symmetric configurations. 
We consider  the particular example of a rotating thin shell that collapses in AdS$_3$, corresponding to a stationary but not static exterior spacetime. This solution will be treated formally in the sense that we do not insist in the physical consistency of the shell's energy momentum tensor. The main interest of this simple exercise is to show that formula (\ref{bdiez}) continues to apply with the same complexity field ${\cal C}_\Sigma^\mu$, despite the existence of `shear' components in the jumping conditions for the Killing vectors. 

The outside spacetime $\mathscr{V}^+$ consists of a rotating BTZ solution (cf. \cite{BTZ})
\bg
 \dd s_+^2 \, = \, -f_+(r)\,  \dd t_+^2 \, +\, \dfrac{ \dd r^2}{f_+(r)} + r^2\left(  \dd \phi_+\, - \, \dfrac{a}{r^2}\, \dd t_+\right)^2\;,
\eg
with blackening factor 
\be
f_+(r) \, =  r^2 - \mu^2\, +\, \dfrac{a^2}{r^2}\;,
\ee
for $a = 4\,GJ$ and $\mu^2 = 8\,GM$ the ADM angular momentum and mass, respectively. We choose the inner spacetime $\mathscr{V}^-$ to be pure AdS$_3$
\bg
 \dd s_-^2 \, = \, -(1+r^2)\,  \dd t_-^2 \, +\, \dfrac{ \dd r^2}{1+r^2} + r^2\,d\phi_-^2\, \;.
\eg

The worldvolume of the shell $\mathcal{W}$ will have metric
\be
 \dd s^2_{\mathcal{W}} = - \dd \tau^2 \,+\,R(\tau)^2\, \dd \psi^2 \;,
\ee
where $\psi$ is a co-rotating angle. Demanding for the continuity of the metric across $\mathcal{W}$ translates then to the set of conditions
\bg
\psi \, = \, \phi_- \, = \, \phi_+ - \,\omega(R)\, t_+\, + \theta(\tau)\;,\\
-1 \, =\, -f_-(R)\, (\dot{t}_-)^2 \, +\, \dfrac{(\dot{R})^2}{f_-(R)}\,=  \, -f_+(R)\, (\dot{t}_+)^2 \, +\, \dfrac{(\dot{R})^2}{f_+(R)}\;,\label{israel1b}
\eg
where the angular frequency of the shell is basically $\omega(R)\, = \, a/R^2$, and the function $\theta(\tau)$ accounts for the variation in the angular frequency of the shell due to its shrinking
\be
\dot{\theta}(\tau)\, =\, \dot\omega(R)\,t_+\;.
\ee

The discontinuity in the extrinsic curvature on $\mathcal{W}$ as seen from $\mathscr{V}^\pm$ will be sourced by the induced energy-momentum tensor of the shell $S_{\mu\nu}$. Since the interior frame is co-rotating with the shell, the situation is the same as for the spherically symmetric collapse in section \ref{sec:collapse}, for which we already know the components of the extrinsic curvature. The calculation from the exterior frame is a little more involved, but it can be done by using the precise form of the outward pointing $\mathcal{W}$-normal  $\left(N_\mathcal{W}\right)_\mu \,=\, \dot{t}_+\, (\text{d}r)_\mu - \dot{R} \,(\text{d}t_+)_\mu$ and velocity field $u^\mu \, = \, \dot{R}\,\partial_r^\mu \, +\, \dot{t}_+\,\partial_{t_+}^\mu \,+\omega\,\dot{t}_+ \,\partial_{\phi_+}^\mu\,$. The second junction conditions can then be expressed as
\bg
S^\tau\,_\tau  = \dfrac{1}{8\pi G}\,\dfrac{\beta_+ -\beta_-}{R}\label{junction2.1}\\
S^\psi\,_\psi = \dfrac{1}{8\pi G}\,\dfrac{\dot{\beta}_+ -\dot{\beta}_-}{\dot{R}}\label{junction2.2}\\
S^\tau\,_\psi =  -\dfrac{1}{8\pi G}\, \omega R\label{junction2.3}
\eg
where $\beta_\pm = \sqrt{\dot{R}^2  + f_\pm(R)}$. 

Let us proceed to calculate the discontinuity in the stationary Killing vector
\be
\Delta \xi ^\mu \, = \, -(\Delta \xi \cdot u) \,u^\mu \, +\, \dfrac{(\Delta \xi\cdot \partial_\psi X )}{R^2}\,\partial_\psi X^\mu\,+\, (\Delta \xi \cdot N_\mathcal{W})\, N_\mathcal{W}^\mu\;.
\ee
It is straightforward to evaluate all these projections, an using \eqref{junction2.1} and \eqref{junction2.3} we can write them as
\be
\Delta \xi ^\mu \, = \, -\left(8\pi G S_{\tau\tau}R \right)\,u^\mu \, -\, \left(8\pi G\,S_{\tau \psi}\,\dfrac{1}{R}\,\right)\partial_\psi X^\mu\,.
\ee

Plugging this result in \eqref{stationarypi}, and noting that the extremal hypersurface $\Sigma$ will in this case intersect $\mathcal{W}$ on a constant $\tau$ circle, we have that the angular discontinuity of the Killing does not contribute to the rate of the complexity since $(N_{\mathcal{W}}\cdot \partial_\psi X)$ vanishes. Moreover, the contribution from the Killing discontinuity in the $u^\mu$ direction has the same form as in the spherically symmetric case, and hence we obtain the same PC duality
\be
{\dot{\cal C}}[{\cal O}_{\rm shell}]\, \, =\,-\int_{\Sigma}\,  N_\Sigma^\mu \;T_{\mu\nu} \;{\cal C}_{\Sigma}^\nu\;,
\ee
where the `complexity field' ${\cal C}_{\Sigma}^\mu \, = -r\,e_y^\mu$. It is tempting to conjecture that the `complexity field' ${\cal C}_{\Sigma}^\mu$ persists to be inward pointing tangent to $\Sigma$ and orthogonal to $\Sigma \cap \mathcal{W}$ for more general situations of thin shells gluing two stationary spacetimes $\mathscr{V}^\pm$ together.

\cleardoublepage

\bibliographystyle{style}
\bibliography{pc.bib}

\end{document}